# On-chip Single Nanoparticle Detection and Sizing by Mode-splitting in an Ultra-high-$Q$ Microresonator

Jiangang Zhu[1], Sahin Kaya Ozdemir[1], Yun-Feng Xiao[1†], Lin Li[2], Lina He[1], Da-Ren Chen[2] & Lan Yang[1]

[1]Department of Electrical and Systems Engineering, [2]Department of Energy, Environmental & Chemical Engineering, Washington University in St. Louis, St. Louis, Missouri 63130, USA.

[†]Present address: State Key Laboratory for artificial microstructure, Institute of Modern Optics, Peking University, Beijing, China

**The ability to detect and size individual nanoparticles with high resolution is crucial to understanding behaviours of single particles and effectively using their strong size-dependent properties to develop innovative products. We report real-time, in-situ detection and sizing of single nanoparticles, down to 30 nm in radius, using mode-splitting in a monolithic ultra-high-$Q$ whispering-gallery-mode (WGM) microtoroid resonator. Particle binding splits a WGM into two spectrally shifted resonance modes, forming a self-referenced detection scheme. This technique provides superior noise suppression and enables extracting accurate size information in a single-shot measurement. Our method requires neither labelling of the particles nor apriori information on their presence in the medium, providing an effective platform to study nanoparticles at single particle resolution.**

With the rapid progress in nanotechnology, a variety of nanoparticles of different materials and sizes have been synthesized and engineered as key components in various applications ranging from solar cell technologies to biomolecular detection (*1-3*). Meanwhile, nanoparticles generated by vehicles and industries have become sources of potential threats to human health and environment (*4, 5*). Techniques such as confocal, near-field optical microscopy, Raman spectroscopy, and fluorescence microscopy, have played central roles in single nanoparticle/molecule detection; however, their widespread use is limited by bulky and expensive instrumentation, long processing time, or the need for labelling (*6-8*). Light scattering techniques, while suitable for label-free detection of individual particles, are hindered by the extremely small scattering cross-sections of single nanoparticles. Detection and sizing of particles with radius R≥40 nm using optical-scattering based particle counters have been reported (*9, 10*). As the interests in nanoparticle applications and awareness of their potential risks to health/environment are increasing, there is a great need to develop new label-free optical techniques to achieve portable, inexpensive and high-resolution devices capable of real-time and in-situ detection of particles surpassing current detection limits.

Attributed to the highly confined microscale mode volume and ultra-high-$Q$, whispering gallery mode (WGM) microresonators (*11, 12*) enable strong light-matter interaction that can be used for ultra-sensitive optical detection. Previously, detection of influenza virus and other bio-molecules have been demonstrated by monitoring frequency shift of a WGM upon binding of targets onto the resonator surface (*13-17*). However, particle detection and sizing based on estimation of the spectral shifts is limited in two ways. First, the shift induced by a nanoparticle is very small, and it is sensitive to laser intensity and frequency fluctuations, thermal noise, detector noise, and environmental disturbance. Thus, discriminating between the interactions of interest and the interfering perturbations becomes difficult. Second, the amount of shift depends on the interaction strength between the particle and the WGM which, in turn, is affected by the location of the particle. Specifically, a small particle with larger overlap with WGM can result in the same shift as a large particle with smaller overlap. This makes size estimation of individual nanoparticles through resonance shifts challenging.

In this study, we develop a different technique to detect and size single nanoparticles by leveraging an interesting physical phenomenon, namely mode-splitting (*18-20*), in a WGM resonator. We demonstrate detection, counting and sizing of individual nanoparticles as small as 30 nm in radius using scattering induced mode-splitting of a WGM in an ultra-high-$Q$ microtoroid (*12*). The demonstrated higher level of sensitivity and resolution can be attributed to: (i) Two standing wave modes (SWM) formed after the adsorption of a particle, share the same resonator and experience the same noise. This allows a self-referencing detection system more immune to noise than the resonance shift based sensing schemes. (ii) The linewidths of the SWMs and the amount of mode splitting allow extracting the accurate size information regardless of where the particle is adsorbed, therefore enabling single-shot size measurement. The approach demonstrated here will assist in realization of on-chip detection and sizing systems with single particle resolution, laying the groundwork to investigate the properties of single particles and their dynamics which cannot be attained using ensemble measurements.

The basis of our study is silica microtoroid resonators possessing two degenerate WGMs with the same resonant frequency and field distributions but opposite propagation directions, i.e. clockwise and counter-clockwise modes. Other resonators, such as microspheres and microdisks also support such degenerate modes. This degeneracy is lifted to split the resonance into a doublet if the resonator deviates from its perfect azimuthal symmetry due to any perturbation in the mode volume, e.g., surface roughness, material inhomogeneity or a scatterer (*18-20*). In this study, we lift the degeneracy by depositing a nanoparticle along the periphery of the microtoroid, where the mode resides.

Our experimental set-up (Fig. 1A) contains a fibre-taper (*21*) coupled microtoroid resonator and a Differential Mobility Analyzer (*22*) (DMA) accompanied with a nozzle for nanoparticle deposition (*23*). The fibre-taper is used to couple light in and out of the microtoroid (Figs. 1A-C). Real-time transmission spectrum is collected by a photodiode connected to an oscilloscope. The $Q$-factors of the resonators used in the experiments are greater than $10^8$. The microtoroids were carefully tested to confirm that there was no intrinsic mode-splitting. We performed experiments using Potassium Chloride (KCl, *n*=1.49) and Polystyrene (PS, *n*=1.59, Thermo Scientific Inc., USA) particles with 30 nm ≤ R≤175 nm. Fig. 1D shows a section of a microtoroid with a KCl nanoparticle.

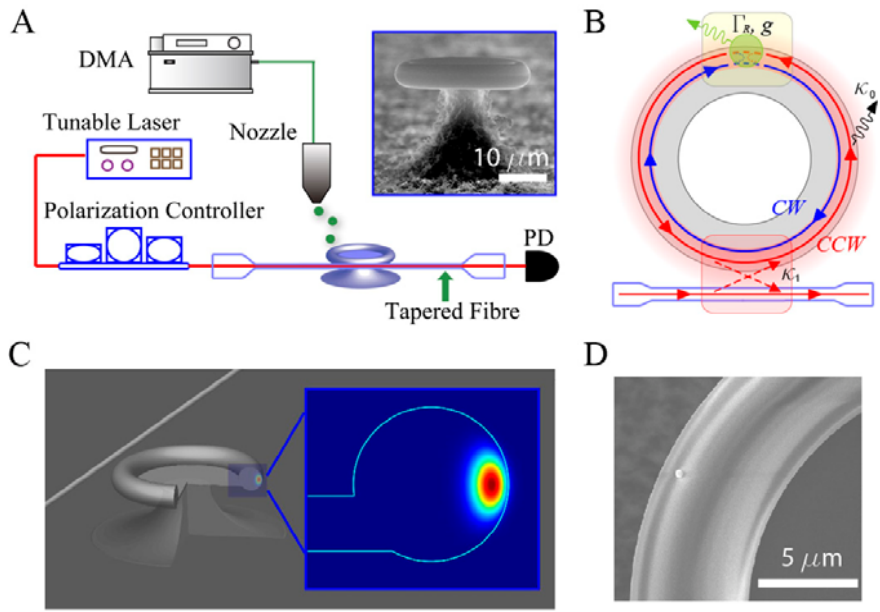

**Fig. 1. (A)** Schematics of the experimental set-up with a Differential Mobility Analyzer (DMA) and a nozzle for particle deposition (23). Light from a tunable laser is coupled in and out of the cavity by a tapered fibre. The transmitted light is monitored by a photodiode (PD). Inset shows a SEM image of a silica microtoroid resonator. **(B)** Illustration of the coupled nanoparticle-microtoroid system. The microtoroid-taper coupling rate is $\kappa_1$, and intrinsic damping rate (material and radiation losses) of the resonator is $\kappa_0$. $g$ and $\Gamma_R$ are the coupling coefficient of the light scattered into the resonator and the additional damping rate due to scattering loss to the environment, respectively. **(C)** Rendering image of fibre coupled microtoroid resonator, and the cross-section of WGM field profile obtained by finite-element-method simulations. **(D)** SEM image of a single nanoparticle with $R$=150 nm deposited on the resonator.

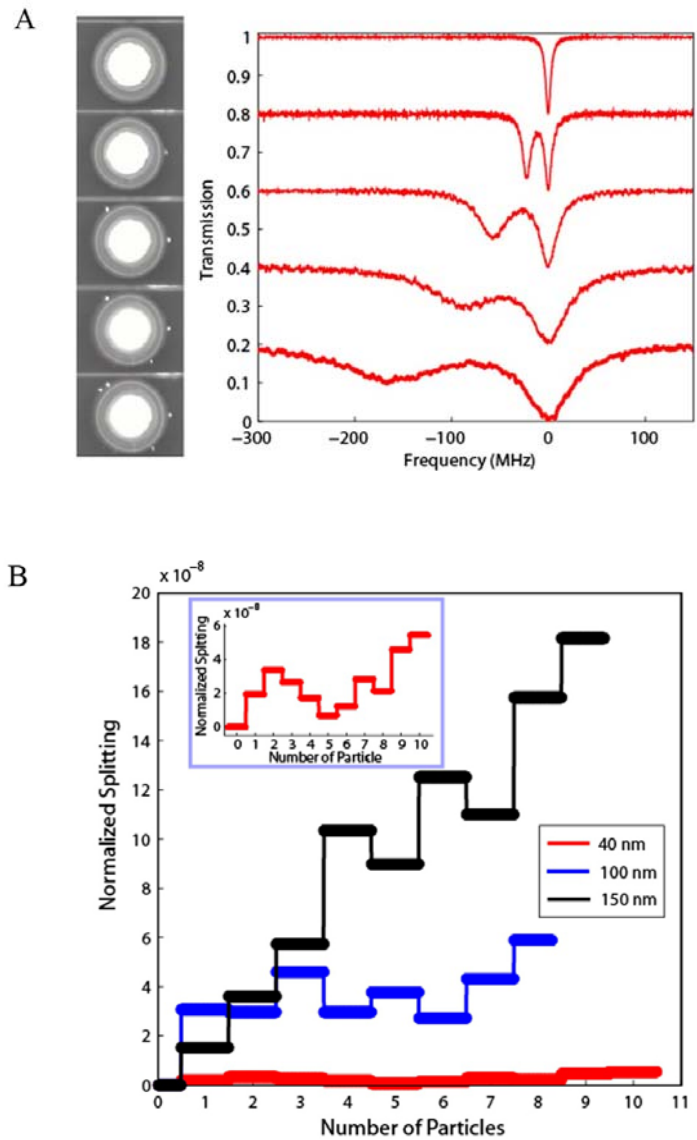

**Fig. 2.** (**A**) Series of normalized transmission spectra taken at 1550 nm wavelength band and the corresponding optical images (assisted by a visible light laser) recorded for four consecutive depositions of KCl nanoparticles on the microtoroid. (**B**) Normalized splitting ($2g/\omega_c$) versus particle number for KCl nanoparticles, where $2g$ denotes amount of splitting and $\omega_c$ corresponds to resonance frequency. Each discrete step (lines are drawn for eye guide) corresponds to a single nanoparticle binding event. Inset shows the enlarged plot for nanoparticles of $R$=40 nm.

Before the particle deposition starts, the transmission spectrum shows a single Lorentzian resonance. After the first particle is deposited, SWMs are formed, which is confirmed by the mode-splitting (double resonances) in the transmission spectra (Fig. 2A). The consecutive particle depositions lead to changes in both the amount of splitting and the linewidths of the resonances. Fig. 2B presents mode splitting versus particle numbers. Discrete steps of various heights are clearly visible indicating that individual nanoparticle adsorption events are resolved. Each adsorbed particle causes redistribution of previously established field; thus the height of each discrete step depends on the positions of the particles relative to the SWMs. Since the particle positions are random, each event does not necessarily lead to enhancement of mode-splitting (*24*).

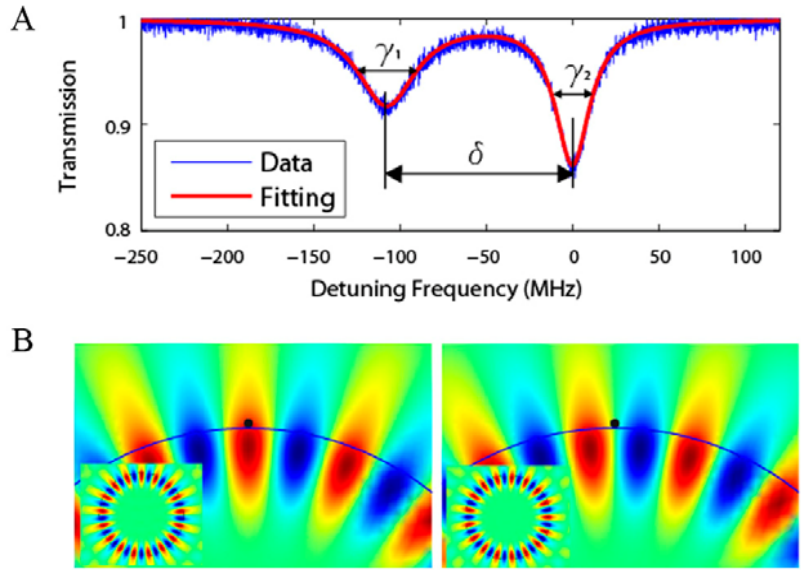

**Fig. 3.** (**A**) Experimentally obtained transmission spectrum (blue) after the deposition of a single nanoparticle and the curve fit (red). A single nanoparticle is detectable provided that $\delta > (\gamma_1 + \gamma_2)/2$ is satisfied. (**B**) Field distribution of symmetric (SM) and asymmetric modes (ASM) relative to the position of the nanoparticle using finite-element-method simulation. Insets show the mode along the periphery of the resonator.

The underlying mechanism responsible for single-particle induced mode-splitting can be intuitively explained as follows (*23*). A nanoparticle in the evanescent field of WGMs acts as a light scatterer. Subsequently, a portion of the scattered light is lost to the environment creating an additional damping channel, while the rest couples back into the resonator and induces coupling between the two counter-propagating WGMs (*18,19*), whose degeneracy is lifted consequently. This creates SWMs that are split in frequency as reflected by the double resonance in the transmission spectrum (Fig. 3A). The SWMs redistribute themselves according to particle location: The symmetric mode (SM) locates the particle at the anti-node while the asymmetric mode (ASM) locates it at the node (Fig. 3B). Consequently, the significantly perturbed SM experiences frequency shift and linewidth broadening. The strength of coupling $g$ is quantified by the doublet-splitting $g = \pi\delta$, where $\delta$ is the detuning of SM from ASM, and the additional linewidth broadening is quantified as $\Gamma_R = \pi|\gamma_1 - \gamma_2|$, where $\gamma_1$ and $\gamma_2$ represent the linewidths of the split modes. In a regime where the particle is considerably smaller than light wavelength $\lambda$, the particle-WGM interaction induces a dipole moment in the particle. This dipole is represented by the polarizability $\alpha = 4\pi R^3(\varepsilon_p - \varepsilon_m)/(\varepsilon_p + 2\varepsilon_m)$ with $\varepsilon_p$ and $\varepsilon_m$ denoting dielectric permittivities of the particle and the medium, respectively. The parameters $g$ and $\Gamma_R$ are given as $g = -\alpha f^2(\mathbf{r})\omega_c/2V_c$ and $\Gamma_R = -g\alpha\omega_c^3/3\pi v^3$ where $\omega_c$ is the angular resonant frequency, $f(\mathbf{r})$ designates normalized mode distribution, $V_c$ is the mode volume, and $v = c/\sqrt{\varepsilon_m}$ with $c$ representing the speed of light. Consequently, we can derive the particle size from $\alpha = -(3\lambda^3/8\pi^2)(\Gamma_R/g)$ where $\Gamma_R$ and $g$ can be measured from the transmission spectrum. Since the value of $\Gamma_R/g$ is independent of the particle position on the resonator, it gives the technique presented here a big advantage over schemes using resonance spectral shift, which is affected by particle positions. If $\varepsilon_s < \varepsilon_m$ ($\varepsilon_s > \varepsilon_m$), SM experiences a red (blue)-shift with respect to ASM. This provides an avenue to differentiate the two situations.

As shown in Fig. 4, our size estimations are in good agreement with the actual sizes of KCl (measured by SEM) and PS (provided by the producer) nanoparticles with estimated mean sizes falling within 2.37% and 1.12% of the actual values on average, respectively. Larger standard deviations for KCl particles are attributed to the non-uniform size and shape of these lab-made particles as verified by SEM. The theoretical lower limit of measurable radius for our device is estimated from $2g > \Gamma_R + (\omega_c/Q)$ (23) as 16 nm for KCl and 15 nm for PS at $\lambda$=670 nm assuming maximum overlap between the particle and WGM. In our experiments, we achieved accurate sizing of PS particles as small as 30 nm in radius in ambient conditions without stabilization of laser power and frequency. Improving the experimental conditions (25) will allow approaching the theoretical limit. The upper limit of size detection is imposed by the condition R ≪ λ. In our experiments, KCl and PS nanoparticles up to R=100 nm were detectable at $\lambda$=670 nm. Bigger particles lead to large linewidth differences which make simultaneous monitoring of split modes difficult (ref). However, those particles could be easily detected and measured at $\lambda$=1450 nm band (23).

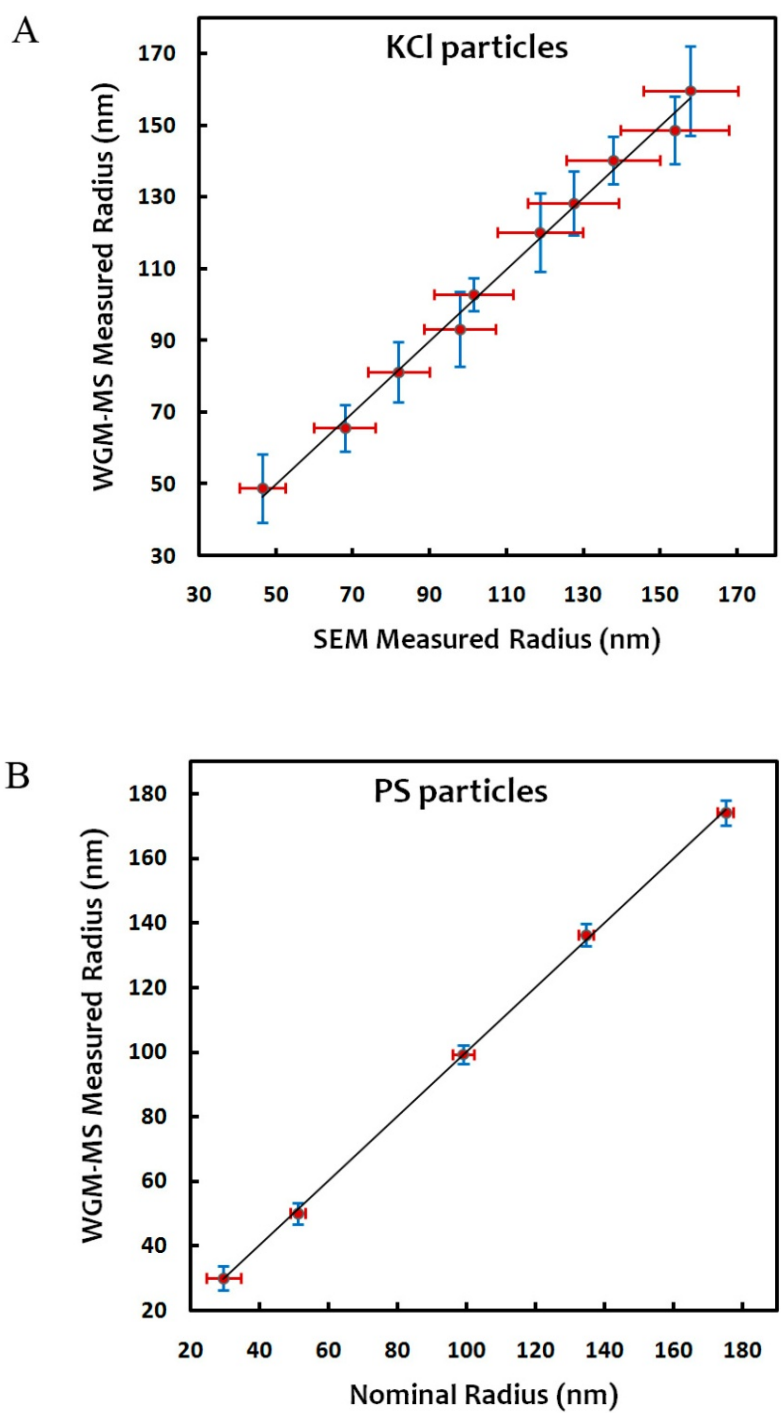

**Fig. 4.** Estimated sizes of particles as a function of their actual sizes are shown for (**A**) KCl particles, and (**B**) PS particles. Error bars denote the standard deviations of the particle sizes.

Size is a key parameter significantly affecting mechanical, optical, electrical, magnetic and biological properties of nanoparticles. It plays a crucial role in the applications of nanoparticles across many scientific disciplines and industries such as biomedicine, opto-electronics, semiconductor processing and environmental science, which will greatly benefit from accurate measurement of individual nanoparticles. Our measurements provide an initial demonstration of detection and sizing of nanoscale dielectric particles using mode-splitting in a monolithic ultra-high-$Q$ microresonator. This highly compact and extremely sensitive detection scheme will allow real-time and in-situ investigating the fundamental properties of nanoparticles, and various parameters (e.g. humidity, temperature) that affects the dynamics of nanoparticles and their interactions (*26, 27*). For example, phase transition studies of atmospheric aerosols (*26*) require real-time monitoring of particles. It usually involves an environmental SEM to mimic the change of humidity and temperature in a chamber, while using our technique these experiments can be performed in ambient conditions. Our device could be configured as a fly-through (*25, 28, 29*) particle detection, sizing and counting sensor. In such a configuration, the particles will fly through the evanescent field of the resonator leading to mode-splitting without being adsorbed. When the particle leaves the sensing volume, the spectrum will recover back as the process is reversible.

In this work we have demonstrated accurate sizing of particles. Moreover, our technique directly reveals the particle polarizability, which depends on particle size, refractive index and geometry. As a result, particles with the same size but different refractive indexes or shapes can be discriminated. Considering that $Q\sim4\times10^8$ and $V_c \sim 1.5 \times 10^{-16}$ m$^3$ have been reported for microtoroids ([11]), we project that the lower detection limit of several nanometers is within reach ([23]). Since $Q$-factor above $10^8$ has been reported for microtoroids in water ([14]), our scheme can be effectively extended to aqueous environments. Therefore, our scheme will enable a new class of sensing and monitoring devices to perform label-free single nanoparticle measurements in various platforms.

# Supplementary materials for:

# "On-chip Single Nanoparticle Detection and Sizing by Mode-splitting in an Ultra-high-$Q$ Microresonator"

Jiangang Zhu[1], Sahin Kaya Ozdemir[1], Yun-Feng Xiao[1†], Lin Li[2], Lina He[1], Da-Ren Chen[2] & Lan Yang[1]

[1]Department of Electrical and Systems Engineering, [2]Department of Energy, Environmental & Chemical Engineering, Washington University in St. Louis, St. Louis, Missouri 63130, USA.

Supporting materials of the experimental set up, procedures and theoretical model is provided for our manuscript (*1*).

## I.     EXPERIMENTAL DETAILS

A schematic illustration of our apparatus for the generation of monodisperse nanoparticles and their deposition on the microtoroid resonator is presented in Fig.1S. The details of the set-up are as follows:

**A. Fabrication of Microtoroid Resonators.** The ultra-high-Q microtoroids used in this work are fabricated from 2 μm thick thermal silica deposited on a silicon wafer. First, series of circular pads of 80 μm in diameter are created through a combination of standard photo-lithography technique and buffered HF etching. Subsequently, these circular pads serve as etch mask for isotropic etching of silicon in $XeF_2$ gas chamber, leaving under-cut silica disks supported by silicon pillar. The silica microdisks are then selectively reflowed using a 30W carbon dioxide ($CO_2$) laser to form a toroidal shape due to surface tension (*2*). The resulting microtoroids have diameters 30-40 μm with measured quality factors Q>$10^8$ in both 665-675 nm and 1430-1480 nm wavelength bands.

**B. Taper-fibre and measurements.** In order to couple light from a tunable laser into and out of the microtoroid resonators, taper-fibres (*3, 4*) were fabricated by pulling single mode fibres on hydrogen ($H_2$) flame. Position of the microtoroid is finely controlled by a piezo stage to adjust the air gap between the taper and microtoroid. During the experiments, two tunable lasers in the 670 nm and 1450 nm wavelength bands were used. Their wavelengths were linearly scanned around the resonance wavelength of the microtoroid. The real-time transmission spectra were obtained by a photodetector followed by an oscilloscope. This enabled a real time monitoring of the transmission spectrum on the oscilloscope. In order to reduce the effects of thermal nonlinearity (*5*) and its distorting effects (*6, 7*) on the high-Q microtoroid due to heat build-up in the resonator, the wavelength scanning speeds of both tunable lasers were set to 40 nm/s, and the laser power was kept around 15 µW.

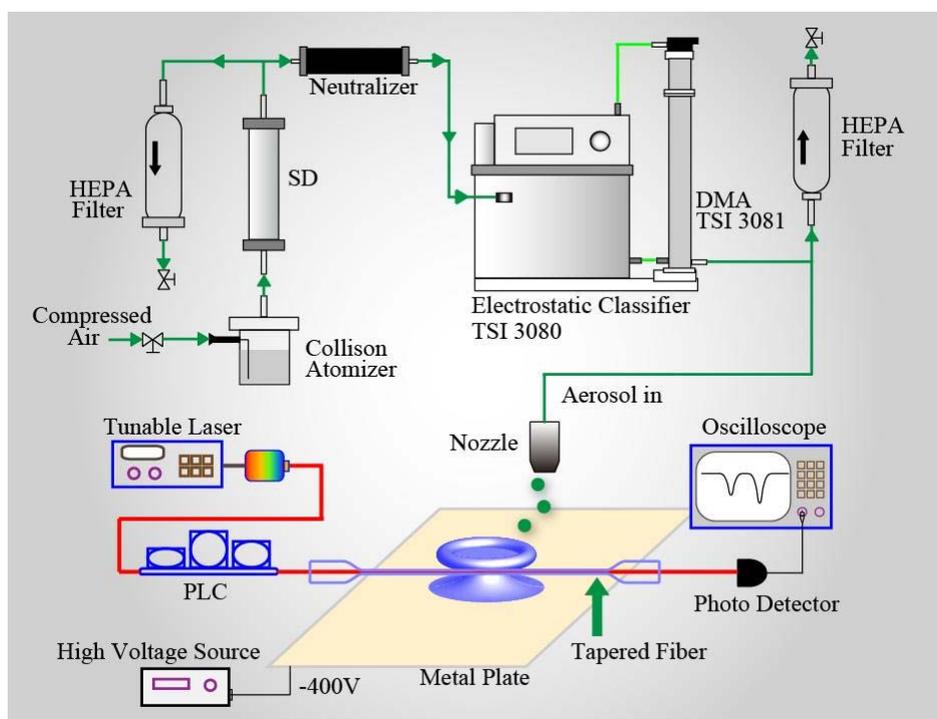

**Fig. S1.** Schematic of the experimental set-up for particle detection. The experimental set-up consists of a differential mobility analyzer (DMA) system for size classification of nanoparticles, a nozzle for depositing nanoparticles onto the microtoroid, and a taper-fibre coupled resonator system. PLC: Polarization controller, PD: Photodetector, SD: Silica gel desiccant dryer, DMA: Differential mobility analyzer.

**C. Differential Mobility Analyzer (DMA) and particle deposition.** We used potassium chloride (KCl) and polystyrene (PS) microspheres (Thermo Scientific, 3000 series, radius 30-175 nm) in mode-splitting and the subsequent particle detection and sizing experiments. The nanoparticles are classified according to their mobility using a DMA (*8*) (Fig. S1). Polydisperse droplets are carried out by compressed air using a Collison atomizer. The solvent in droplets is then evaporated in the dryer with the silica gel desiccant. KCl solid particles are further neutralized by a $Po^{210}$ radioactive source such that they have a well-defined charge distribution. Particles are sent to a DMA where they are classified according to their electrical mobility. Thus, particles within a narrow range of mobility can exit through the output slit of the DMA. The flow rate is controlled and the ratio of particle flow rate to the sheath flow (particle-free air flow) rate was set to 1:10. The resulting monodisperse particle flow has a concentration of about $10^5/cm^3$ and a geometrical standard size deviation of approximate 5% (*9*).

A nozzle with a tip inner diameter of 80 µm was placed at about 150 µm above the microtoroid to deliver the nanoparticles to its mode volume. In order to force the particle's trajectory towards the microtoroid, a metal sheet connected to a -400V source was placed under the silicon chip to exert an electrical field on the particle. When a particle reaches the microtoroid and adsorbed on its surface, the interaction of the WGM with the particle and re-distribution of the resonator field due to the scattering from the particle leads to a mode-splitting which could be observed in the transmission spectrum. The particle-microtoroid binding events are described by a poisson process. In the experiments, we set the particle concentration very low so that the average time interval between two events was longer than 15 seconds. This allowed us to shut the aerosol flow immediately after a mode splitting was observed in the transmission spectrum resulting one and only one particle adsorbed in the mode volume of the microtoroid. The particles deposited outside the mode volume on the microtoroid do not affect the WGM, so they have no effect on the resonance spectrum. To investigate the bonding between the nanoparticle and microtoroid, we kept the microtoroids with adsorbed nanoparticles in a gel-box for two weeks and didn't see any noticeable change in the resonance spectrum.

This suggests that particle attachment to the microtoroid is stable under ambient conditions.

## II.   THEORETICAL MODEL

To understand our experimental observations and to extract the relevant information on the nanoparticle from the measured transmission spectra, we re-visited the theoretical model developed in Refs. (*10, 11*).

**A.   Nanoparticle Induced Mode-Splitting in WGM Microtoroid Resonator.** A perfect azimuthally symmetric microresonator supports two counter-propagating WGMs (clockwise: CW and counter-clockwise: CCW) with the degenerate resonant frequency $\omega_c$ and the same field distribution function $f(\mathbf{r})$. Defining the annihilation (creation) operators $a_{cw}$ ($a_{cw}^\dagger$) and $a_{ccw}$ ($a_{ccw}^\dagger$) for the counter-propagating WGMs, and $b_j$ ($b_j^\dagger$) for the *j*-th reservoir mode with frequency $\omega_j$ , the free Hamiltonian of the resonator-reservoir system is written as

$$H_0 = \hbar\omega_c\big(a_{cw}^\dagger a_{cw} + a_{ccw}^\dagger a_{ccw}\big) + \sum_j \hbar\omega_j b_j^\dagger b_j \tag{S1}$$

In the presence of a scatterer, one of the modes, say CW couples to the scatterer. The scattered light will couple-back to either the CW or the CCW mode. The same is true when the CCW couples to the scatterer. Assuming the same coefficient *g* for all of these coupling processes, and denoting the coupling of the CW and CCW modes to the reservoir modes with the same coupling coefficient of $g'$, we can write the coupling Hamiltonians as

$$H_1 = \hbar g\big(a_{cw}^\dagger a_{cw} + a_{cw}^\dagger a_{ccw} + a_{ccw}^\dagger a_{cw} + a_{ccw}^\dagger a_{ccw}\big) \tag{S2}$$

$$H_2 = \sum_j \hbar g'\big(a_{cw}^\dagger b_j + a_{cw} b_j^\dagger + a_{ccw}^\dagger b_j + a_{ccw} b_j^\dagger\big) \tag{S3}$$

Then the Heisenberg equation of motion for the coupled system is

$$\frac{da_{cw}}{dt} = \frac{1}{i\hbar}[a_{cw}, H] - \frac{\kappa_0 + \kappa_1}{2}a_{cw} - \sqrt{\kappa_1}a_{cw}^{in} \tag{S4}$$

$$\frac{da_{ccw}}{dt} = \frac{1}{i\hbar}[a_{ccw}, H] - \frac{\kappa_0 + \kappa_1}{2}a_{ccw} - \sqrt{\kappa_1}a_{ccw}^{in} \tag{S5}$$

$$\frac{db_j}{dt} = \frac{1}{i\hbar}\left[b_j, H\right] \tag{S6}$$

where $\kappa_0$ and $\kappa_1$ denote the intrinsic damping and the fibre taper-resonator coupling rates, respectively, $a_{cw}^{in}$ and $a_{ccw}^{in}$ correspond to the input CW and CCW fields, and $H = H_0 + H_1 + H_2$. Using the bosonic commutation relations, Eqs. (S4-S6) can be expressed as

$$\frac{da_{cw}}{dt} = -i\left[(\omega_c + g)a_{cw} + g a_{ccw} + \sum_j g' b_j\right] - \frac{\kappa_0 + \kappa_1}{2}a_{cw} - \sqrt{\kappa_1}a_{cw}^{in} \tag{S7}$$

$$\frac{da_{ccw}}{dt} = -i\left[(\omega_c + g)a_{ccw} + g a_{cw} + \sum_j g' b_j\right] - \frac{\kappa_0 + \kappa_1}{2}a_{ccw} - \sqrt{\kappa_1}a_{ccw}^{in} \tag{S8}$$

$$\frac{db_j}{dt} = -i\left[\omega_j b_j + g'(a_{cw} + a_{ccw})\right] \tag{S9}$$

For a subwavelength scatterer (Rayleigh scatterer), the interaction between the WGM and the scatterer can be modeled using the dipole approximation (*12*) where a dipole in the scatterer is induced by the electric field of the coupled WGM. Then in the case of elastic Rayleigh scattering where a photon in the incident mode is scattered into the j-th mode of the reservoir, the coupling coefficients are given as

$$g = -\frac{\alpha f_c^2(\boldsymbol{r})\omega_c}{2V_c}, \quad g' = -\frac{\alpha f_c^2(\boldsymbol{r})\omega_c}{2\sqrt{V_c V_j}}(\hat{n}_m \cdot \hat{n}_j) \tag{S10}$$

where $f_c^2(\boldsymbol{r})$ accounts for the cavity mode functions of the CW and CCW modes, $V_c$ and $V_j$ denote the quantization volumes of the WGM and the vacuum modes, respectively, and $\hat{n}_m$ and $\hat{n}_j$ are the unit vectors of the fields. In Eq. (S10), $\alpha$ is the polarizability of the scatterer which for a spherical scatterer of radius $R$ can be expressed as $\alpha = 4\pi R^3(\varepsilon_p - \varepsilon_m)/(\varepsilon_p + 2\varepsilon_m)$ where $\varepsilon_p$ and $\varepsilon_m$ denote electric permittivities of the particle (scatterer) and the surrounding medium, respectively. Damping rates due to coupling to the reservoir via Rayleigh scattering can be derived using Weisskopf-Wigner approximation from Eq. (S9) as

$$\Gamma_R = \frac{\alpha^2 f_c^2(\boldsymbol{r})\omega_c^4}{6\pi\nu^3 V_c} \tag{S11}$$

where $\nu = c/\sqrt{\varepsilon_m}$, and $c$ is the speed of light in vacuum. Then Eqs. (S7) and (S8) can be simplified into

$$\frac{da_{cw}}{dt} = -i[(\omega_c + g) + \frac{\Gamma_R + \kappa_0 + \kappa_1}{2}]a_{cw} - (ig + \frac{\Gamma_R}{2})a_{ccw} - \sqrt{\kappa_1}a_{cw}^{in} \qquad (S12)$$

$$\frac{da_{ccw}}{dt} = -i[(\omega_c + g) + \frac{\Gamma_R + \kappa_0 + \kappa_1}{2}]a_{ccw} - (ig + \frac{\Gamma_R}{2})a_{cw} - \sqrt{\kappa_1}a_{ccw}^{in} \qquad (S13)$$

Defining the normal modes of the resonator as $a_{\pm} = (a_{cw} \pm a_{ccw})/\sqrt{2}$ and that of the input modes as $a_{\pm}^{in} = (a_{cw}^{in} \pm a_{ccw}^{in})/\sqrt{2}$, we find that in the steady-state regime normal modes can be expressed as

$$\left[-i(\Delta - 2g) + \frac{\Gamma + \Gamma_R}{2}\right]a_+ + \sqrt{\kappa_1}a_+^{in} = 0 \qquad (S14)$$

$$\left(-i\Delta + \frac{\kappa_0 + \kappa_1}{2}\right)a_- + \sqrt{\kappa_1}a_-^{in} = 0 \qquad (S15)$$

where $\Gamma = \Gamma_R + \kappa_0 + \kappa_1$, and $\Delta = \omega - \omega_c$ denotes the laser-cavity detuning. It is clear that the symmetric standing mode ('+') has a detuning $2g$ from the degenerate WGM, and its damping rate is $2\Gamma_R + \kappa_0 + \kappa_1$ while the asymmetric standing mode ('-') is not affected by the scatterer. In the absence of CCW input, i.e., $a_{ccw}^{in} = 0$, we find the transmission and reflection coefficient of the coupled system as

$$t = 1 - \frac{\kappa_1 \beta}{\beta^2 - (ig + \Gamma_R/2)^2}, \qquad r = \frac{\kappa_1(ig + \Gamma_R/2)}{\beta^2 - (ig + \Gamma_R/2)^2} \qquad (S16)$$

where $\beta = -i\Delta + ig + \Gamma/2$, and we used $a_{cw}^{out} = a_{cw}^{in} + \sqrt{\kappa_1}a_{cw}$ as the input-output relation of the fiber-taper coupled resonator system.

**B. Particle Size Estimation.** The mode-splitting observed in the transmission spectrum can be utilized to estimate the size of Rayleigh scatterer. Assuming that the surrounding medium is air, from Eqs. (S10, S11) we find $\Gamma_R/g = 8\pi^2\alpha/3\lambda^3$ where $\lambda$ is the resonance wavelength before splitting. Then $\alpha = 3(\Gamma_R/g)\lambda^3/8\pi^2$, which implies that polarizability of the particle can be calculated accurately from the measured values of $g$ and $\Gamma_R$. Subsequently, nanoparticle size can be accurately estimated provided that its refractive index is known. Fig. S2A shows a typical transmission spectrum with a double Lorentzian resonance captured by an oscilloscope. We extract $g$ and $\Gamma_R$ by fitting a double Lorentzian function expressed as

$$f(\omega) = 1 - \frac{A_1 \gamma_1^2/4}{(\omega - \omega_1)^2 + \gamma_1^2/4} + 1 - \frac{A_2 \gamma_2^2/4}{(\omega - \omega_2)^2 + \gamma_2^2/4} \tag{S17}$$

to the acquired transmission spectrum. In Eq. (S17), $\omega_1$, $\omega_2$ denote the locations of resonance dips, $\gamma_1$, $\gamma_2$ designate the linewidths of the resonances, and $A_1$ and $A_2$ correspond to depths of the resonances. Fig. S2 shows the acquired data and the fitted double Lorentzian curve.

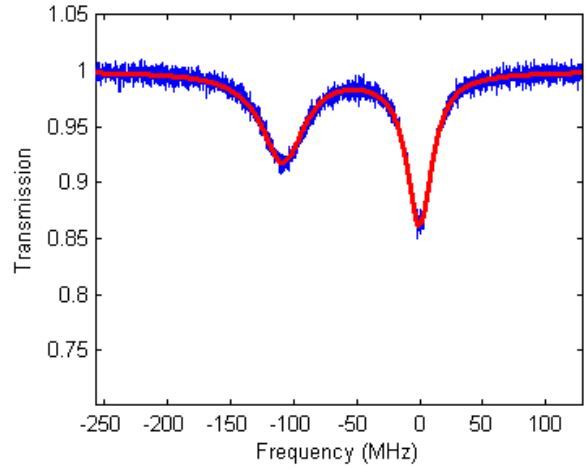

**Fig. S2.** Typical experimental transmission spectra and the Lorentzian curve fit after deposition of a single KCl nanoparticle of nominal radius R=75 nm.

During the fitting process, the parameters in Eq. (S17) are varied until the best fit is obtained by minimizing mean square error. The amount of splitting then is calculated from $\delta = |\omega_1 - \omega_2|$ and equated to the coupling coefficient as $2g = \delta$. The values of $\gamma_1$ and $\gamma_2$ are used to obtain the additional damping parameter using $2\Gamma_R = \gamma_1 - \gamma_2$. Then we find

$$\alpha = 4\pi R^3 \frac{n_p^2 - 1}{n_p^2 + 2} = \frac{3\lambda^3}{8\pi^2} \cdot \frac{\gamma_1 - \gamma_2}{\delta} \tag{S18}$$

where $n_p$ denotes the refractive index of the particle. Consequently the particle radius $R$ is given by

$$R = \left[ \frac{(3\lambda^3/8\pi^2) \cdot (\gamma_1 - \gamma_2)/\delta}{4\pi(n_p^2 - 1)/(n_p^2 + 2)} \right]^{\frac{1}{3}} \tag{S19}$$

**C. Detection limits.** A nanoparticle is detectable when the splitting is distinguishable. This requires that the amount of splitting quantified as $|2g|$ should be greater than the sum of the frequency linewidth $\omega_c / Q$ of the WGM and the additional damping rate $2\Gamma_R$. Then the lower detection limit can be derived from

$$|2g| = \frac{\alpha f_c^2(\boldsymbol{r})\omega_c}{V_c} > \frac{\omega_c}{Q} + \Gamma_R, \tag{S20}$$

where we used $\kappa_0 + \kappa_1 = \omega_c/Q$. Since $\Gamma_R \ll \omega_c/Q$, for the sake of simplicity we can ignore $\Gamma_R$ arriving at

$$\alpha > \frac{1}{f_c^2(\boldsymbol{r}) \cdot Q/V_c}. \tag{S21}$$

Consequently the lower limit of detectable particle radius is found as

$$R = \left[\frac{1}{4\pi(n_p^2 - 1)/(n_p^2 + 2)} \frac{1}{f_c^2(\boldsymbol{r}) \cdot Q/V_c}\right]^{\frac{1}{3}} \tag{S22}$$

where $Q/V_c$ is the Purcell factor.

The highest $Q$ achieved for microtoroid is around $4 \times 10^8$, and the microtoroid should not be smaller than 30 μm in major diameter (Fig. S3A) to maintain this value as further decrease of radius will increase radiation loss (*13*). Such a microtoroid yields a mode volume of $V_c = 1.47 \times 10^{-16}$ m$^3$ given by numerical simulation using finite-element method. Figs S3B, C show the simulated WGM field distribution in the cross-section of the toroid ring quantified as $f_c(\boldsymbol{r})$ around 670 nm wavelength. On the surface of this microtoroid, the maximum value of $f_c(\boldsymbol{r})$ is 0.36. Inserting the values of $V_c$, $Q$ and $f_c(\boldsymbol{r})$ in Eq. (S22), we calculate the lower limit as $R=9.2$ nm and $R=8.7$ nm for KCl ($n_p=1.49$) and PS ($n_p=1.59$) nanoparticles, respectively.

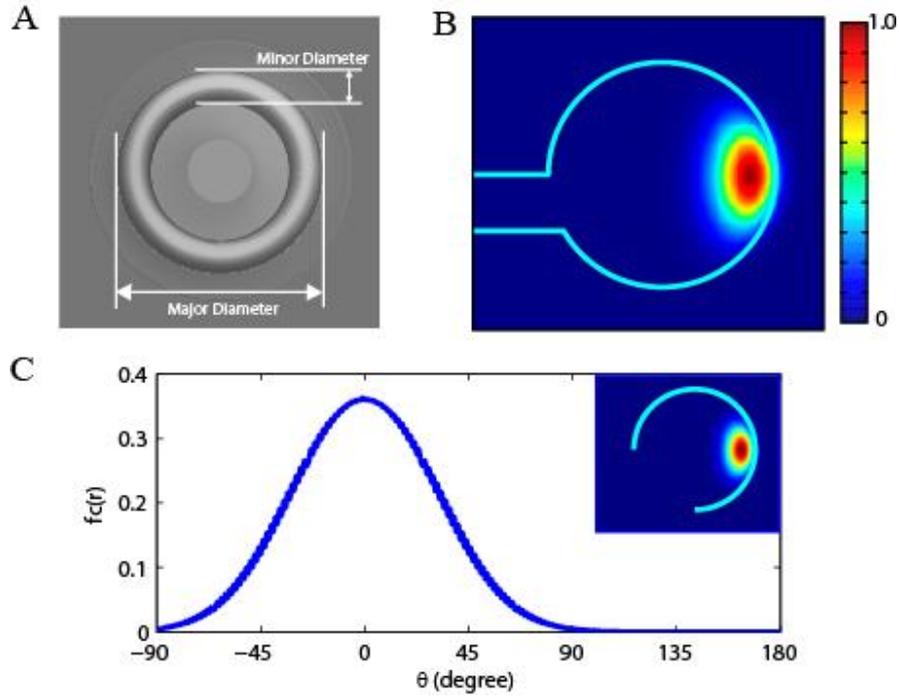

**Fig. S3.** **(A)** Illustration of a toroid showing size notations. **(B)** Normalized WGM field distribution in the cross-section of a microtoroid (Major diameter: 30 μm, and minor diameter: 5 μm) obtained by simulation for light wavelength of 670 nm. **(C)** The normalized field strength $f_c(r)$ along the outer surface of the microtoroid cross-section shown in **(B)**. The inset shows the trajectory of data points.

The upper detection limit, on the other hand, can be estimated from the conditions of Rayleigh scattering and dipole approximation which assume that $R \ll \lambda$. In our experiments, KCl and PS nanoparticles up to 100 nm in radius are detectable in 670 nm band, and the spectrum agrees well with the theoretical prediction derived using dipole approximation. For particles above this size, we see a large additional damping as $\Gamma_R \propto R^6$ (see Eq. S11), consequently the $Q$-factor of the symmetric mode (SM) becomes very low. The big difference in $Q$-factors of SM and ASM resonances makes it very difficult to monitor them simultaneously (Fig. S4A). However, when the wavelength is switched to near-infrared band (1450 nm) both resonances become clear (Fig. S4B) as the damping rate which scales as $\Gamma_R \propto 1/\lambda^4$ significantly decreases.

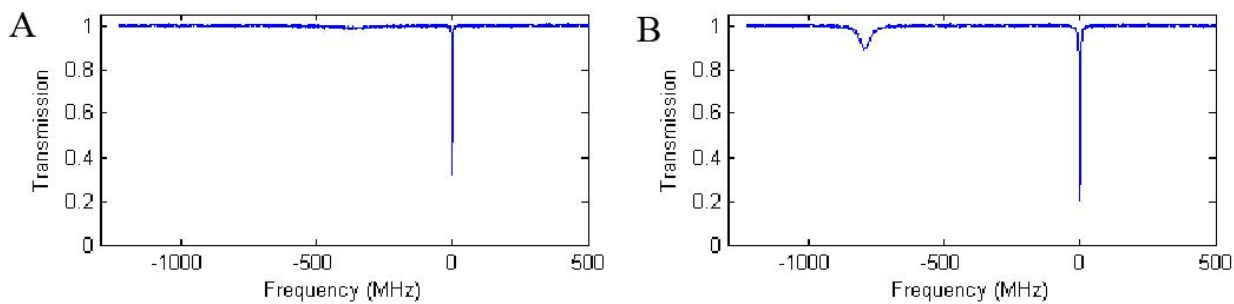

**Fig. S4.** Transmission spectra obtained at two different wavelengths for a single PS nanoparticle of R=110 nm. **(A)** For visible light at 670 nm band, symmetric mode cannot be clearly observed. **(B)** For near-infrared light at 1450 nm band, the symmetric mode is seen due to the decreased damping at this wavelength band.